\documentclass[11pt]{article}
\usepackage[dvips]{graphicx}
\usepackage{amssymb}
\usepackage[usenames]{color}
\begin{document}
\begin{center}
{\LARGE\bf The Consistent Result of Cosmological Constant From
Quantum Cosmology and Inflation with Born-Infeld Scalar Field }
 \vskip 0.15 in
 $^\dag$H.Q.Lu$^1$, $^\ddag$W.Fang$^1$, Z.G.Huang$^2$, P.Y.Ji$^1$\\
$^1$Department~of~Physics,~Shanghai~University,\\
~Shanghai,~200444,~P.R.China\\
$^2$Department of Mathematics and Physics,
\\Huaihai Institute of Technology, Lianyungang, 222005, P.R.China
\footnotetext{$\dag$ Alberthq$_-$lu@staff.shu.edu.cn}
\footnotetext{$\ddag$ xiaoweifang$_-$ren@shu.edu.cn }
 \vskip 0.5 in
\centerline{\bf Abstract} \vskip 0.2 in
\begin{minipage}{5.5in} { \hspace*{15pt}\small The Quantum cosmology with
Born-Infeld(B-I) type scalar field is considered.  In the extreme
limits of small cosmological scale factor the wave function of the
universe can also be obtained by applying the methods developed by
Hartle-Hawking(H-H) and Vilenkin. H-H wave function predicts that
most Probable cosmological constant $\Lambda$ equals to
$\frac{1}{\eta}$($\frac{1}{2\eta}$ equals to the maximum of the
kinetic energy of scalar field). It is different from the original
results($\Lambda=0$) in cosmological constant obtained by
Hartle-Hawking. The Vilenkin wave function predicts a nucleating
unverse with largest possible cosmological constant and it is larger
than $1/\eta$.  The conclusions have been nicely to reconcile with
cosmic inflation. We investigate the inflation model with B-I type
scalar field, and find that $\eta$ depends on the amplitude of
tensor perturbation $\delta_h$, with the form $\frac{1}{\eta}\simeq
\frac{m^2}{12\pi[(\frac{9\delta_{\Phi}^2}{N \delta_h^2})^2-1]}.$ The
vacuum energy in inflation epoch depends on the tensor-to-scalar
ratio $\frac{\delta_h}{\delta_{\Phi}}$. The amplitude of the tensor
perturbation ${\delta_{h}}$ can, in principle, be large enough to be
discovered. However, it is only on the border of detectability in
future experiments. If it has been observed in future, this is very
interesting to determine the vacuum energy in inflation epoch.

}
 { \hspace*{15pt}\small \\ {\bf Keywords:} Cosmological Constant; Quantum Cosmology; Inflation;
 Born-Infeld Type Scalar Field
{\bf PACS:} 98.80.Cq, 04.65.+e, 11.25.-w}
\end{minipage}
\end{center}
\newpage
\section{Introduction}
 \hspace{15pt}
Astronomical observations indicate that the cosmological constant is
not zero and it has the same order of magnitude as matter energy
density(density parameter $\Omega_{\Lambda}$ of the vacuum energy
$\sim$ 0.73). \par Before 1998 year a crude experimental upper bound
on $\Lambda$ or vacuum energy density $\rho_V$ is provided as
$\rho_V\lesssim 10^{-29}g/cm^3\sim10^{-47}GeV^4$[1]. However when
the universe approximates to planck scale the energy difference
between the symmetry and broken symmetry phase of vacuum is $10^{18}
GeV$. The effective vacuum energy density$(10^{18}GeV)^4$ exceeds
observational limit by some 120 orders of magnitude. There are many
number of symmetries which seem to be broken in the present
universe, including chiral symmetry, electroweak symmetry and
possibly supersymmetry. Each of these would give a contribution to
$\rho_V$ that would exceed the upper limit by at least forty orders
of magnitude. It is very difficult to believe that the cosmological
constant is fine tune so that after all the symmetry breakings, the
effective vacuum energy density satisfies the upper bound. What one
would like to is some mechanism by which the cosmological constant
$\Lambda$ could relax to near zero. Weinberg has described five
approaches to find such a mechanism, including Anthropic
considerations, superstrings and supersymmetry, Adjustment
Mechanisms, changing Gravity, Quantum Cosmology. At present, the
approaches which based on quantum cosmology is most promising[1]. In
1984 Hawking described how in quantum cosmology there could arise a
distribution of values for the cosmological constant. Hawking
introduces a 3-form gauge field $A_{\mu\nu\eta}$ or scalar field
$\phi$. According to the general ideas of Euclidean quantum
cosmology, he obtained that probability density is proportional to
$e^{3\pi/G\Lambda}$. The probability density has an infinite peak
for $\Lambda\rightarrow0_+$. The most probable cosmological constant
will be those with very small values[2]. Coleman considers the
effect of topological fixtures known as wormholes. He argued that if
wormholes exist, they have the effect of making the cosmological
constant vanish[3]. However these conclusions have been hard to
reconcile with cosmic inflation[4]
\par Recently, the problem of cosmological constant based on quantum
cosmology has been investigated by many authors. Kalinin and
Melnikov discuss quantization of closed isotropic cosmological model
with a cosmological constant which is realized by the
Wheeler-Dewitt(WDW) equation. It is shown that such quantization
leads to interesting results, in particular, to a finite lifetime of
the system, and appearance of the universe as penetration via the
barrier. These purely quantum effects appear when cosmological
constant is larger than zero[5]. Capozziello and Gavattini have
defined the cosmological constant is an eigenvalue of WDW equation
with f(R) theories of gravity. The explicit calculation is performed
for a schwarzschild metric where one-loop energy is derived by the
zeta function regularization method and renormalized running
cosmological constant is obtained [6]. There are papers on the
quantum cosmology by Lemos and Monerat et.al[7]. The quantization of
the Friedmann-Robertson-Walker spacetime in the presence of negative
cosmological constant was used in the papers. They have concluded
that there are solutions which avoid singularities(big bang, big
crunch) at the quantum level[7]. Moguigan have discussed a seesaw
mechanism of cosmological constant in the context of quantum
cosmology[8].  The FRW quantum cosmology in the non-Abelian
Born-Infeld theory has been discussed by Moniz[9]. Reference [10]
have discussed Quantum birth of the universe in gravitational theory
of varying cosmological constant.
\par In the 1930's Born and Infeld[20] attempted to eliminate the
divergent self energy of the electron by modifying Maxwell's theory.
Born-Infeld electrodynamics follows from the lagrangian
$L_{BI}=b^2(\sqrt{1-(1/2b^2)F_{\mu\nu}F^{\mu\nu}}-1)$, Where
$F_{\mu\nu}$ is the electromagnetic field tensor.  Our B-I type
lagrangian has been first proposed by Heisenberg in order to
describe the process of meson multiple production connected with
strong field regime[21]. The Born-Infeld type action also appears in
string theory[22]. In the important paper[23] it was demonstrated
that the leading-order term in the expansion in $\partial F$ of the
condition of conformal invariance of the open string sigma model
follows indeed from the BI action. Static and spherically symmetric
solutions of the B-I type scalar field have been recently
investigated qualitatively by Oliveiva[24]. Furthermore our B-I type
scalar field lagrangian is the special case of tachyon largrangian
$u(\varphi)\sqrt{1-g^{\mu\nu}\varphi_{,\mu}\varphi_{,\nu}}$ when the
potential $u(\varphi)=constant$. V.Mukhanov and A.Vikman[18] have
investigated inflation with an analogous scalar field lagrangian
$\alpha^2(\sqrt{1+\frac{g^{\mu\nu}\varphi_{,\mu}\varphi_{,\nu}}{\alpha^2}}-1)-V(\varphi)$.
The speed of cosmological perturbations $c_s^2>1$, which is
different with our B-I scalar field lagrangian(see Eq(34)). We have
investigated quantum cosmology and dark energy model in our B-I type
scalar field[11].
\par In this paper, we combine inflation with quantum cosmology in
the B-I type scalar field and obtain interesting results that the
vacuum energy(corresponding cosmological constant) in inflation
epoch depends on the tensor-to-scalar ratio
$\frac{\delta_h}{\delta_{\Phi}} $.
\par The present paper is organized as follows. In section 2 we consider quantum cosmology with B-I type scalar
 field, obtain the wheeler-Dewitt(W-D) equation of our B-I scalar
 field model. In section 3 we apply  Hartle-Hawking's
 method to obtain the wave function of the universe. The Vilenkin's quantum tunneling approach is also considered.
  The probability density obtained from Hartle-Hawking method
 is proportional to $e^{3\pi/G(\Lambda-\frac{1}{\eta})}$. The
 probability density has an infinite peak for
 $\Lambda\rightarrow\frac{1}{\eta}$(the maximum kinetic energy
 $\frac{1}{2}\dot\phi^2$ is $\frac{1}{2\eta}$). i.e, the vacuum energy equals to two time of maximum kinetic energy of scalar field.
 The Vilenkin wave function predicts a nucleating
unverse with largest possible cosmological constant and it is larger
than $1/\eta$. In section 4 we have discussed the inflation with B-I
type scalar field. We find that $\eta$ depends on the amplitude of
tensor perturbation $\delta_h$, with the form $\frac{1}{\eta}\simeq
m^2/12\pi[(\frac{9\delta_{\Phi}^2}{N \delta_h^2})^2-1]$. We conclude
our results in the last Section.
 \section{WD Equation with B-I type Scalar Field}
 The action of the gravitational field interacting with a
 Born-Infeld type scalar field is given by
 \begin{equation}S=\int\frac{R}{4\pi G}\sqrt{-g}d^4x+\int
 L_s\sqrt{-g}d^4x\end{equation}
 where we have chosen units so that $c=1$, R is the Ricci Scalar
 curvature and the lagrangian $L_s$ of the B-I scalar field[11] is
 \begin{equation}L_s=\frac{1}{\eta}(1-\sqrt{1-\eta g^{\mu\nu}\varphi_{,\mu}\varphi_{,\nu}})-V(\varphi) \end{equation}
Where $\eta$ is a constant and $V(\varphi)$ is the potential of
vacuum field. We shall consider that the potential of the vacuum
field $V(\varphi)$ is a constant. There from a physical point of
view, it is equivalent to cosmological constant.
\par At the planck time, the quantum effects played the main role in
the universe. So it is suitable to describe the dynamics and
evolution of very early universe by using the cosmological wave
function $\Psi(h_{ij}, \varphi)$ defined on the superspace of all
three-metrics $h_{ij}$ and material fields $\varphi$. In the
superspace it satisfies the Wheeler-DeWitt(WD) equation:
\begin{equation}
\hat{H}\Psi=0\end{equation} $\hat{H}$ is a second-order differential
operator in the superspace. In principle, $\Psi(h_{ij},\varphi)$
should contain the answer  to all meaningful questions one can ask
about the evolution of the very early universe. In order to find out
the solution of the WD equation, we shall apply the minisuperspace
model---a Robertson-Walker(RW) space-time metric. The B-I type
scalar field is given by Eq.(2). In the minisuperspace there are
only two degrees of freedom: $a(t)$ and $\varphi(t)$. The RW
space-time metric is
\begin{equation}ds^2=-dt^2+a^2(t)[\frac{dr^2}{1+kr^2}+r^2(d{\theta}^2+sin^2\theta d{\varphi}^2)]\end{equation}
Using Eq.(2) and by integrating with respect to space-components the
action (1) becomes(the upper-dot means the derivative with respect
to the time t):
\begin{equation}S=\int\frac{3\pi}{4G}(1-\dot a^2)adt+\int2\pi^2a^3[\frac{1}{\eta}(1-\sqrt{1-\eta\dot\varphi^2})-V]dt=\int \mathcal{L}_gdt+\int \mathcal{L}_sdt\end{equation}
From the Eule-lagrange equation
\begin{equation}\frac{d}{dt}(\frac{\partial \mathcal{L}_s}{\partial
\dot\varphi})-\frac{\partial \mathcal{L}_s}{\partial \varphi}=0
\end{equation} We can obtain
\begin{equation}\dot\varphi=\frac{c}{\sqrt{a^6+\eta c^2}} \end{equation}
where $c$ is integral constant. From the above equation we know that
cosmological scale factor $a(t)$ is very large or small when
$\dot\varphi$ is very small or large respectively. The maximum of
kinetic energy is $1/2\eta$ from Eq.(7).
\par To quantize the
model, we first find out the canonical momenta $P_a=\partial
{\mathcal{L}}_g/\partial\dot{a}=-(3/2G)a\dot{a}$,
$P_\varphi=\partial
{\mathcal{L}}_\varphi/\partial\dot{\varphi}=2\pi^2a^3\dot{\varphi}/\sqrt{1-\eta\dot{\varphi}^2}$
and the Hamiltonian
$H=P_a\dot{a}+P_\varphi\dot{\varphi}-{\mathcal{L}}_g-{\mathcal{L}}_s$.
$H$ can be written as the follows
\begin{equation}H=-\frac{G}{3\pi a}P_a^2-\frac{3\pi}{4G}a[1-\frac{8\pi G}{3}a^2 V(\varphi)]-\frac{2\pi^2a^3}{\eta}[1-\sqrt{1+\frac{\eta P^2_\varphi}{4\pi^4a^6}}]\end{equation}
For $\dot{\varphi}^2\ll\frac{1}{\eta}$, the Hamiltonian Eq.(8) can
be simplified by using the Taylor expansion, and the terms smaller
than $\dot{\varphi}^6$ can be ignored, so the Hamiltonian becomes
\begin{equation}H=-\frac{G}{3\pi a}P_a^2-\frac{3\pi}{4G}a[1-\frac{8\pi G}{3}a^2 V(\varphi)]+\frac{P_\varphi^2}{4\pi^2a^3}-\frac{\eta P_\varphi^4}{64\pi^6a^9}\end{equation}
If $\dot{\varphi}$ is very large(${\dot\varphi}^2\sim1/\eta$),
Eq.(8) becomes
\begin{equation}H=-\frac{G}{3\pi a}P_a^2-\frac{3\pi}{4G}a\{1-\frac{8\pi G}{3}a^2[ V(\varphi)-\frac{1}{\eta}]\}\end{equation}
The WD equation is obtained from $\hat{H}\psi=0$, Eqs.(9) and (10)
by replacing $P_a\rightarrow-i(\partial/\partial a)$ and
$P_\varphi\rightarrow i(\partial/\partial\varphi)$. Then we obtain
\begin{equation}[\frac{\partial^2}{\partial a^2}+\frac{p}{a}\frac{\partial}{\partial a}-\frac{1}{a^2}\frac{\partial^2}{\partial\tilde{\Phi}^2}-\frac{\eta}{16\pi^4a^8}\frac{\partial^4}{\partial\tilde{\Phi}^4}-U(a,\tilde{\Phi})]\psi=0\end{equation}
and
\begin{equation}[\frac{\partial^2}{\partial a^2}+\frac{p}{a}\frac{\partial}{\partial a}-u(a,\tilde{\Phi})]\psi=0\end{equation}
where $\tilde{\Phi}^2=4\pi G\varphi^2/3$ and the parameter $p$
represent the ambiguity in the ordering of factor $a$ and
$\partial/\partial a$ in the first term of Eqs.(9) and (10). The
variation of $p$ does not affect the solution of Eqs.(9) and (10).
In following discussion, we shall set $p=-1$[12,16]. We have also
denoted
\begin{equation}U(a,\tilde{\Phi})=(\frac{3\pi}{2G})^2a^2[1-\frac{8\pi G}{3}a^2V(\tilde{\Phi})]\end{equation}
\begin{equation}u(a,\tilde{\Phi})=(\frac{3\pi}{2G})^2a^2\{1-\frac{8\pi G}{3}a^2[V(\tilde{\Phi})-\frac{1}{\eta}]\}\end{equation}
Eqs.(11) and (12) are the WD equations corresponding to the action
(1) in the case of large and small scale factor $a$ respectively. If
scale factor $a$ is very large, the genral solution of Eq.(1) is
given by
\begin{equation}\psi(a, \varphi)\sim \frac{a}{a_0}Z_{\frac{1}{3}}(\frac{2\tilde{\mu} a^3}{3})e^{-k\varphi}\end{equation}
Where $a_0$ is the planck length, $k$ is an arbitrary constant. and
$\tilde{\mu}=2\pi^2V(\varphi)$. The $\psi(a, \varphi)$ is an
oscillatory function with respect to scale factor $a$[11].
\section{The Vilenkin and Hartle-Hawking Method}
\par Next we will use Vilenkin's quantum tunneling[12] approach to
consider the cosmology in case of very large
$\dot{\varphi}$(correspondingly very small $a(t)$). Eq.(12) has the
form of a one-dimensional Schr\"{o}dinger equation for a "particle"
described by a coordinate $a(t)$, which is zero energy and moves in
a potential $u$. The classically allowed region is $u\leq0$ or
$a\geq H^{-1}$, with $H=[\frac{8\pi G}{3}(V-\frac{1}{\eta})]^{1/2}$
. In this region, disregarding the pre-exponential factor, the WKB
solutions of Eq.(12) are
\begin{equation}\psi_{\pm}^{(1)}(a)=exp\{[\pm i\int_{H^{-1}}^aP(a')da']\mp\frac{i\pi}{4}\}\end{equation}
The under-barrier($a<H^{-1}$, classically forbidden or Euclidean
region) solutions are
\begin{equation}\psi_{\pm}^{(2)}(a)=exp\{[\pm\int^{H^{-1}}_a|P(a')|da']\}\end{equation}
where $P(a)\equiv\sqrt{|-u(a)|}$.
\\The classical momentum conjugate to $a$ is $P_a=-a\dot{a}$. For
$a>H^{-1}$, we have
\begin{equation}(-i\frac{d}{da})\psi^{(1)}_{\pm}(a)=\pm P(a)\psi^{(1)}_{\pm}(a)\end{equation}
and thus $\psi^{(1)}_-(a)$ and $\psi^{(1)}_+(a)$ describe the
expanding and contracting universe respectively. The tunneling
boundary condition requires that only the expanding component should
be present at large $a$,
\begin{equation}\psi_T(a>H^{-1})=\psi_-^{(1)}(a)\end{equation}
The under-barrier wave function is found from WKB connection formula
\begin{equation}\psi_T(a<H^{-1})=\psi_+^{(2)}(a)-\frac{i}{2}\psi^{(2)}_-(a)\end{equation}
The growing exponential $\psi_-^{(2)}(a)$ and the decreasing
exponential $\psi_+^{(2)}(a)$ have comparable amplitudes at the
nucleation point $a=H^{-1}$, but away from that point the decreasing
exponential dominates
\begin{equation}\psi_T(a<H^{-1})\approx \psi_+^{(2)}(a)=exp[\frac{\pi}{2GH^2}(1-H^2a^2)^{\frac{3}{2}}]\end{equation}
The "tunneling amplitude" is proportional to
\begin{equation}\frac{\psi_T{(H^{-1})}}{\psi_T(0)}=e^{-\frac{\pi}{2GH^2}}\end{equation}
The corresponding probability density
\begin{equation}P_T\propto e^{\frac{-\pi}{GH^2}}
\end{equation} From Eq.(23) we obtain the result
that the tunneling wave function predicts a nucleating universe with
the largest possible vacuum energy(i.e, the largest possible
cosmological constant), but cosmological $\Lambda$ must be larger
than $1/\eta$. In other word, the vacuum energy is larger than two
time of the maximum of kinetic energy of scalar field. It is correct
condition for the inflation.  Eqs.(21,23) can be obtain by an
alternative method, devised by Zeldovich and starobinsky[13],
Rubakov[14] and Linde[15]. The Eq.(23) predicts that a typical
initial value of the field $\varphi$ is given by $V(\varphi)\sim
M_p^4$(if one does not speculate about the possibility that
$V(\varphi)\gg M_p^4$), which leads to a very long stage of
inflation.
\par The
Hartle-Hawking(H-H) no boundary wave function is given by the path
integral[16]
\begin{equation}\psi_{HH}=\int[dg][d\varphi]e^{-S_E(g,\varphi)}\end{equation}
In order to determine $\psi_{HH}$, we assume that the dominant
contribution to the path integral is given by the stationary points
of the action(the instantons) and evaluates $\psi_{HH}$ simply as
$\psi_{HH}\sim e^{-S_E|_{saddle-point}}$. When $\dot\varphi$ is very
large, $\dot\varphi^2\sim 1/\eta$, from action (5) we can obtain
\begin{equation}S=\int\frac{3\pi}{4G}[(1-\dot
a^2)a]dt-\int2\pi^2a^3(V-\frac{1}{\eta})dt\end{equation} The
corresponding Euclidean action $S_E=-i(S)_{continue}$ is
\begin{equation}
S_E=\int\frac{3\pi}{4G}[1+(\frac{da}{d\tau})^2]ad\tau+\frac{3\pi}{4G}\int
a^3H^2d\tau\end{equation} Where $H^2=\frac{8\pi
G}{3}(V-\frac{1}{\eta})$ and $\tau=it$. From action (25), we can
also obtain that the $a(t)$ satisfying the following classical
equation of motion
\begin{equation}-(\frac{da}{dt})^2-1+H^2a^2=0\end{equation}
The solution of Eq.(27) is the de-Sitter space with
$a(t)=H^{-1}cosh(Ht)$. The corresponding Euclidean version(replacing
$t\rightarrow -i\tau$) of Eq.(27) is
\begin{equation}(\frac{da}{d\tau})^2-1+H^2a^2=0\end{equation}
The solution of Eq.(28) is
\begin{equation}a(\tau)=H^{-1}sin(H\tau)\end{equation}
 We consider a saddle point approximation to the path integral (24), use Eqs.(26,29), and obtain
 \begin{equation}\psi_{HH}(a)\propto exp[-\frac{\pi}{2GH^2}(1-H^2a^2)^{\frac{3}{2}}]\end{equation}
The only one difference between the H-H's wave function(30) and
Vilenkin's wave function(21) is the sign of the exponential factor.
The H-H's wave function(30) gives out the probability density
\begin{equation}P_{HH}\propto e^{\frac{\pi}{GH^2}}\end{equation}
The H-H's probability density (31) is the same as Vilenkin's one
(23), except a sign in the exponential factor. The probability
density(31) is peaked at $V-(1/\eta)=0$(note, here $H^2=\frac{8\pi
G}{3}(V-1/\eta))$ and it predicts a very possible universe with a
positive cosmological constant $\frac{1}{\eta}$. The corresponding
vacuum energy equals to two time of maximum of kinetic energy of
scalar field. It predicts a correct condition for inflation
cosmology. It is different from previous result predicted by Hawking
that cosmological constant equals zero[2].
\section{The
Inflation with B-I type scalar field}
 \par The lagrangian of B-I
type scalar field
\begin{equation}L=K(X)-V=\frac{1}{\eta}[1-\sqrt{1-2\eta
X}]-\frac{1}{2}m^2{\phi}^2\end{equation} We assume that for the
homogenous scalar field $X=\frac{1}{2}{\dot\phi}^2$. Here the $L$
plays the role of pressure $p$. The corresponding energy density
\begin{equation}\rho=2XL_{,X}-L=\frac{1}{\eta}[(1-2\eta X)^{-\frac{1}{2}}-1]+\frac{1}{2}m^2{\phi}^2\end{equation}
The effective speed of sound (i.e, speed of propagation of the
cosmological perturbation) is
\begin{equation}c_s^2=\frac{p_{,X}}{\rho_{,X}}=1-2\eta X\end{equation}
 The stability condition
with respect to the high frequency cosmological perturbation
requires $c_s^2>0$. We can find that $c_s^2<1$ and when
$\eta\rightarrow 0$, $c_s^2=1$. In the nonlinear scalar field model
considered by Mukhanov and Vikman[18], the speed of sound $c_s^2>1$.
Let us consider a R-W space-time with small perturbations:
\begin{equation}ds^2=(1+2\Phi)dt^2-a(t)^2[(1-2\Phi)\delta_{ik}+h_{ik}]dx^idx^k\end{equation}
Where $\Phi$ is the gravitational potential characterizing scalar
metric perturbations and $h_{ij}$ is a traceless, transverse
perturbations describing the gravitational waves. The equation of
Einstein gravitation field:
\begin{equation}H^2+\frac{k}{a^2}=\frac{8\pi}{3}\rho\end{equation}
where $H^2=(\frac{\dot a}{a})^2$, the dot denotes the derivative
with respect to time $t$. In inflation epoch, the term $k/a^2$ in
(36) becomes negligibly small compared with $H^2$. \par The
effective energy-moment conservation law is
\begin{equation}\dot\rho+3H(\rho+p)=0\end{equation}
We obtain the following equation for scalar field by varying action
of B-I type scalar field.
\begin{equation}\ddot\phi+3c_s^2H\dot\phi+\frac{V_{,\phi}}{\rho_{,X}}=0\end{equation}
From the Eqs.(36,38) of the field motion it is clear that if
following slow-roll conditions
\begin{equation}XK_{,X}\ll V,~~~and~~~ K\ll V, ~~~|\ddot\phi|\ll\frac{V_{,\phi}}{\rho_{,X}}\end{equation}
and satisfied for at least 75 e-folds then we have a successful
slow-roll inflation due to the potential $V(\varphi)$. Considering
the canonical scalar field with $K=X$, one can take a flat potential
$V(\phi)$ so that $X\ll V$(for more than 75 e-folds). It is the
standard slow-roll inflation[17] and in this case $c_s=1$. In
contrast to ordinary slow-roll inflation one can have any speed of
sound and $c_s<1$. It is important to the amplitude of the final
scalar perturbations(during the postinflationary,
radiation-dominated epoch) and the ratio of tensor to scalar
amplitudes on supercurvature scales are given by[18]
\begin{equation}\delta_{\Phi}^2\simeq\frac{64}{81}[\frac{\rho}{c_s(1+p/\rho)}]_{c_sk\simeq Ha}\end{equation}
\begin{equation}\frac{\delta_{h}^2}{\delta_{\Phi}^2}\simeq27[c_s(1+p/\rho)]_{k\simeq Ha}\end{equation}
Here it is worthwhile reminding that all physical quantities on the
right hand side of Eqs.(40,41) have to be calculated  during
inflation at the moment when perturbations with wave number $k$
cross corresponding horizon:$c_s k\simeq Ha$ for $\delta_{\Phi}$ and
$k\simeq Ha$ for $\delta_h$ respectively. The amplitude of the
scalar perturbations $\delta_{\Phi}$ is a free parameter of the
inflationary theory which is taken to fit the
observations($10^{-5}$).
\par In the slow-roll regime, Eqs(36,38) reduce to
\begin{equation}H\simeq \sqrt{\frac{4\pi}{3}}m\varphi\end{equation}
\begin{equation}3p_{,X}H\dot\varphi+m^2\varphi\simeq 0\end{equation}
From Eq.(42,43) we can obtain a slow-roll solution
$\dot\varphi\simeq-\frac{mc_I}{\sqrt{12\pi}}$, then we obtain
\begin{equation}\varphi=\varphi_0-\frac{mc_I}{\sqrt{12\pi}}t\end{equation}
Where $c_I=(1+\frac{\eta m^2}{12\pi})^{-1/2}$, it is the sound speed
during inflation. The sound speed is smaller than the speed of
light, approaching it as $\eta\rightarrow0$. The effective energy
density and pressure are given by
\begin{equation}\rho=\frac{1}{\eta}[\frac{1}{c_I}-1]+\frac{1}{2}m^2
\varphi^2,~~p=\frac{1}{\eta}[1-c_I]-\frac{1}{2}m^2\varphi^2\end{equation}
To determine $a(\varphi)$ we use
$\dot\varphi\simeq-\frac{mc_I}{\sqrt{12\pi}}$ to rewrite the
equation (42) as
\begin{equation}-\frac{mc_I}{\sqrt{12\pi}}\frac{dlna}{d\varphi}\simeq\sqrt{\frac{4\pi}{3}}m\varphi\end{equation}
and obtain
\begin{equation}a(\varphi)\simeq a_f exp[\frac{2\pi}{c_I}(\varphi_f^2-\varphi^2)]\end{equation}
Where $a_f$ and $\varphi_f$ are the values of the scale factor and
scalar field at the end of inflation. The inflation is over when
$(\rho+p)/\rho\simeq c_I/(6\pi)^{1/2}$ becomes of order unity, that
is, at $\varphi\sim\varphi_f=(c_I/6\pi)^{1/2}$. After that the field
$\varphi$ begins to oscillate and decays. Given a number of e-folds
before the end of inflation N, we find that at this time
$2\pi\varphi^2/c_I\sim N$, hence $(\rho+p)/\rho\simeq1/3N$ does not
depend on $c_I$. Thus, for a given scale, which cross the Hubble
scale N e-folds before the end of inflation, the tensor-to-scalar
ratio is[18]
\begin{equation}\frac{\delta_{h}^2}{\delta_{\Phi}^2}\simeq27[c_I(1+p/\rho)]\simeq\frac{9c_I}{N}\end{equation}
Next one can estimate the mass m which is necessary to produce the
observed $\delta_{\Phi}\sim10^{-5}$. Using
$(2\pi\varphi^2/c_I)\simeq N$, $(\rho+p)/\rho\simeq1/3N $ and from
Eq.(40) one can obtain $m\simeq3\sqrt{3\pi}\delta_{\Phi}/4N$. Then
one can obtain $m\sim 2.4\times10^{-7}$ for $N\sim 75$. It is
similar to the usual chaotic inflation[19]. The spectral index of
scalar perturbations is

\begin{equation}n_s-1\simeq-3(1+\frac{p}{\rho})-H^{-1}\frac{d[ln(1+\frac{p}{\rho})]}{dt}=-\frac{2}{N}\end{equation}
This is exactly the same tilt as for the usual chaotic inflation.
Finally we obtain by Eq.(48) and $c_I=(1+\frac{\eta
m^2}{12\pi})^{-\frac{1}{2}}$
\begin{equation}\frac{1}{\eta}\simeq \frac{m^2}{12\pi[(\frac{9\delta_{\Phi}^2}{N \delta_h^2})^2-1]}\end{equation}
\begin{center}\vspace{0.5cm}
\includegraphics[angle=270,width=10cm]{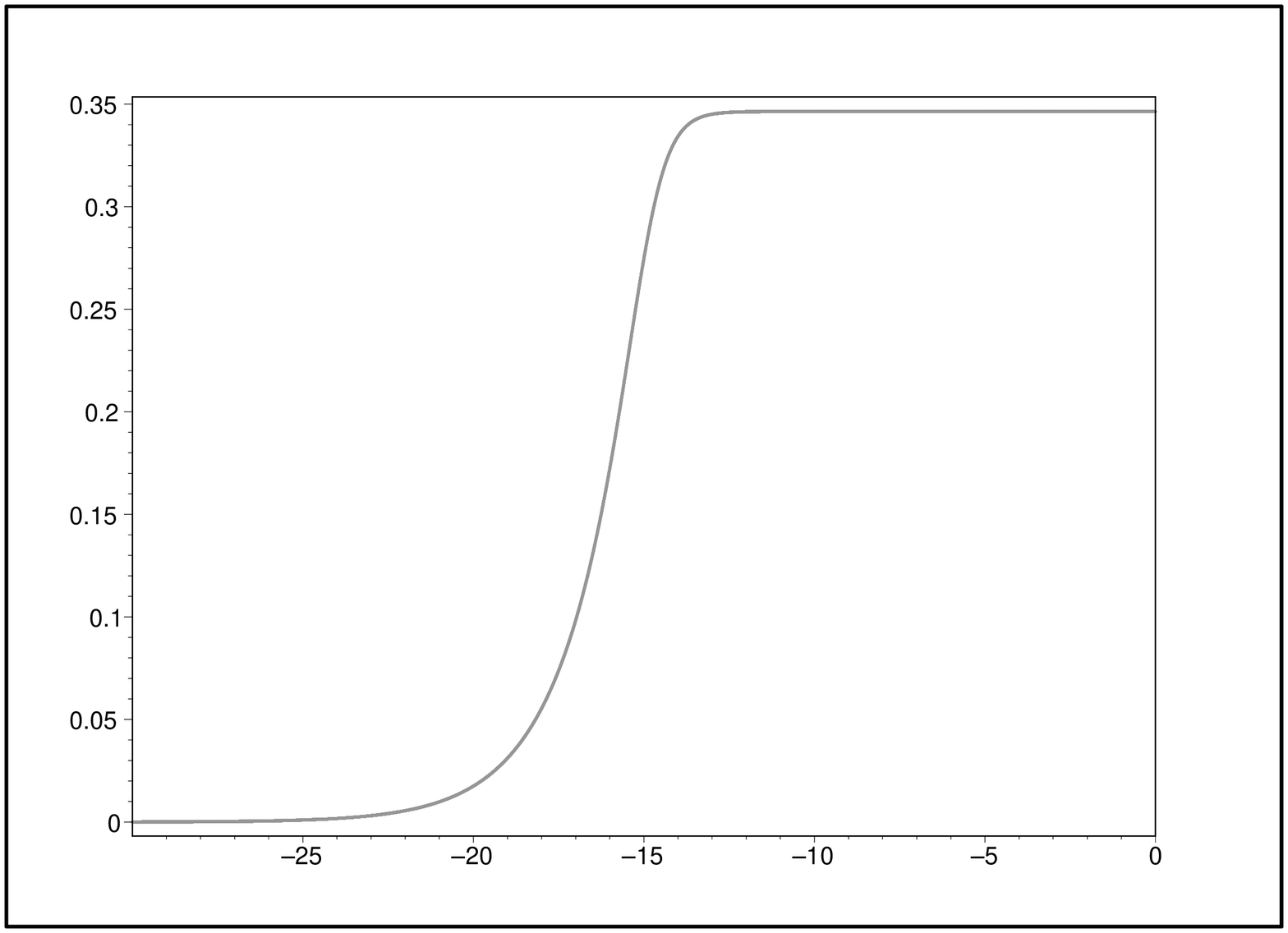} \end{center}\hfill
\begin{minipage}{5.5in} Fig1: The value of $\frac{\delta_h}{\delta_{\Phi}}$ with $\frac{1}{\eta}$. The horizontal axis represents $log_{10}\frac{1}{\eta}$
  and the vertical axis represents
  $\frac{\delta_h}{\delta_{\Phi}}$\end{minipage} \\
\par We list the details in TABLE 1
:\\ \\
\begin{tabular}{|c|c|c|c|c|c|c|c|}
\hline
$\frac{1}{\eta}$ & 1& $10^{-3}$ & $10^{-5}$ & $10^{-7}$ & $10^{-9}$ & $10^{-11}$ \\
\hline $\frac{\delta_h}{\delta_{\Phi}}$ &0.3464101614& 0.3464101614&0.3464101614&0.3464101602&0.3464100291&0.3463969309\\
\hline
$\frac{1}{\eta}$ & $10^{-13}$ & $10^{-15}$ & $10^{-17}$ & $10^{-19}$ & $10^{-21}$ & $10^{-23}$ \\
\hline $\frac{\delta_h}{\delta_{\Phi}}$ &0.3450994646&0.2747269024&0.09836927818& 0.03115736170&0.009852982485&0.003115787144\\
\hline
\end{tabular} \\ \\
\par From our Vilenkin and Hartle-Hawking wavefunction(Eqs.(21,30) we obtain that $\Lambda\geq\frac{1}{\eta}$. In inflation
epoch, the evolution of the field $\varphi$ is very slow, so that
this field acts only as a cosmological constant
$\Lambda(\varphi)=\frac{8\pi V(\varphi)}{M_p^2},$ where $M_p$ is the
plank mass[19]. Noted that $\Lambda \geq \frac{1}{\eta}=1, 10^{-3},
\cdots, 10^{-23}$ correspond to the vacuum energy $\rho_V\geq
\frac{M_p^4}{8\pi}, \frac{10^{-3}M_p^4}{8\pi}, \cdots,
\frac{10^{-23}M_p^4}{8\pi}$. From the Fig.1 or Table 1 we can
conclude that
 only when the vacuum energy $\rho_V<\frac{10^{-14}M_p^4}{8\pi}$ the
 value of $\frac{\delta_h}{\delta_{\Phi}}$ begins decrease rapidly.
 When $\rho_V\rightarrow 0$, $\frac{\delta_h}{\delta_{\Phi}}\rightarrow
 0$. The tensor perturbations can be seen indirectly in the B-mode of the CMB
polarization. The amplitude of the tensor perturbations can, in
principle, be large enough to be observed. However, it is only on
the border of detectability in future experiments. If it has been
observed in future, this is very interesting to define the
cosmological constant.
\section{Conclusion}
Weinberg has described five directions that have been taken in
trying to solve the problem of the cosmological constant. The five
approaches respectively are Anthropic considerations, superstrings
and supersymmetry, Adjustment Mechanisms, changing Gravity, Quantum
Cosmology. At present, all of the five approaches to the
cosmological constant problem remain interesting . The approach
which based on quantum cosmology is most promising[1]. In quantum
cosmology with B-I type scalar field, our Hartle-Hawking wave
function predicts that the most probable value of cosmological
constant is $\frac{1}{\eta}$ while our Vilenkin wavefunction
predicts $\Lambda>\frac{1}{\eta}$. These results are correct
condition for inflation cosmology. The parameter $\eta$ can be
obtained by the tensor perturbation from our inflation model. The
tensor perturbations can be seen indirectly in the B-mode of the CMB
polarization. The amplitude of the tensor perturbations can, in
principle, be large enough to be observed. However, it is only on
the border of detectability in future experiments. If it has been
observed in future, this is very interesting to determine the
cosmological constant in the inflation epoch.

\section{Acknowledgement}
 \hspace*{15 pt}This work is partly supported by National Nature Science
  Foundation of China and by Shanghai Municipal Science and Technology Commission No.04dz05905. \\

{\noindent\Large \bf References} \small{
\begin{description}
\item {1.} {S.Weinberg, Rev.Mod.Phys\textbf{61}, 1(1989).}
\item {2.} {S.W.Hawking, Phys.Lett.B\textbf{134}, 403(1984b).}
\item {3.} {S.Coleman, Nucl.Phys.B\textbf{307}, 867(1988a);\\
            S.Coleman, Harvard University Preprint No. HUTP-88/A022.}
\item {4.} {P.J.Steinhardt and N.Turok, astro-ph/0605173.}
\item {5.} {M.I.Kalinin and V.N.Melnikov, Grav.Cosmology\textbf{9}, 227(2003).}
\item {6.} {S.Capozziello and R.Garattini, Class.Quant.Grav.\textbf{24}, 1627(2007).}
\item {7.} {N.A.Lemos, G.A.Monerat, E.V.Correa Silva, G.Oliveira-Neto and L.G.Ferreira Filho, Phys.Rev.D\textbf{73}, 044022(2006);\\
            N.A.Lemos, G.A.Monerat, E.V.Correa Silva, G.Oliveira-Neto and L.G.Ferreira Filho, gr-qc/0702119(accepted by PRD).}
\item {8.} {M.McGuigan, hep-th/0602112.}
\item {9.} {P.V.Moniz, Class.Quantum Grav.\textbf{19}, L127-L134(2002).}
\item {10.} {T.Harko, H.Q.Lu and M.K.Mak, Europhys.Lett.\textbf{49}, 814(2000).}
\item {11.} {W.Fang, H.Q.Lu and Z.G.Huang, Class.Quantum Grav\textbf{24},3799-3811(2007);\\
             W.Fang, H.Q.Lu and Z.G.Huang, Int.J.Mod.Phy.A\textbf{22},2173-2195(2007);\\
             W.Fang, H.Q.Lu, B.Li and K.F.Zhang, Int.J.Mod.Phys.D\textbf{15},1947-1961(2006);\\
             W.Fang, H.Q.Lu, Z.G.Huang and K.F.Zhang, Int.J.Mod.Phys.D\textbf{15},199(2006);\\
             K.F.Zhang, W.Fang and H.Q.Lu, Int.J.Theor.Phys\textbf{46}1341-1358(2006);\\
             H.Q.Lu, Int.J.Mod.Phys.D\textbf{14}, 355(2005);\\
             H.Q.Lu T.Harko and K.S.Cheng, Int.J.Mod.Phys.D\textbf{8}, 2625(1999).}
\item {12.} {A.Vilenkin, Phys.Rev.D\textbf{33}, 3560(1986);\\
            A.Vilenkin, Phys.Rev.D\textbf{37}, 888(1988);\\
            A.Vilenkin, Phys.Rev.D\textbf{50}, 2581(1994).}
\item {13.} {Ya.B.Zeldovich and A.A.Starobinsky, Sov.Astron.Lett\textbf{10}, 135(1984).}
\item {14.} {V.A.Rubakov, Phys.Lett.\textbf{148B}, 280(1984).}
\item {15.} {A.D.Linde, JETP\textbf{60}(1984), 211; Lett.NuovoCim\textbf{39}, 401(1984); Phys.Lett.\textbf{129B}, 177(1983); Phys.Lett.\textbf{116B}, 335(1982)}
\item {16.} {J.B.Hartle and S.W.Hawking, Phys.Rev.D\textbf{28}, 2960(1983);\\
            S.W.Hawking, Phys.Rev.D\textbf{32}, 2489(1985);\\
            S.W.Hawking, Nucl.Phys.B\textbf{239}, 257(1984);\\
            S.W.Hawking and D.N.Page, Nucl.Phys.B\textbf{264}, 185(1986).}
\item {17.} {A.Riotto, hep-ph/0210162.}
\item {18.} {V.Mukhanov, {\it "Physical Foundations of Cosmology"}, 2005, Cambridge Univ.Press;\\
            V.Mukhanov and A.Vikman, JCAP\textbf{0602}, 004(2006);\\
            A.Vikman, astro-ph/0606033.}
\item {19.} {A.Linde, Physica Scripta T\textbf{36}, 30(1991).}
\item {20.} {M.Born and L.Infeld, Proc.Roy.Soc\textbf{144},425(1934).}
\item {21.} {W.Heisenberg,Z.Phys.\textbf{133},79(1952);\textbf{126},519(1949);\textbf{113},61(1939).}
\item {22.} {J.Polchinski, String Theory Vol.1 Cambridge University Press(1998).}
\item {23.} {A.Abouelsaood, C.G.Callan, C.R.Nappi and S.A.Yost, Nuclear Physics \textbf{B280}[FS18],599-624(1987).}
\item {24.} {H.P.de Oliveira,J.Math.Phys.\textbf{36}, 2988(1995).}
\end{description}
\end{document}